\documentclass[reprint,aps,prstab,showpacs,showkeys,longbibliography]{revtex4-1}
\usepackage{amsmath,amsfonts,graphicx,amssymb,array,natbib, textcase, bm}
\usepackage{hyperref,amsbsy,subcaption,url}

\begin{document}

\title{Influence of the electron beam emittance on the polarization of a laser--electron X-ray generator}

\author{R.M. Feshchenko},
\email{rusl@sci.lebedev.ru}
\author{A.V. Vinogradov}
\author{I.A. Artyukov}
\affiliation{P.N. Lebedev Physical Institute of RAS, 53 Leninski Pr., Moscow, Russia, 119991}
\date{\today}

\begin{abstract}
In this paper we analyze the polarization of the X-ray radiation coming from laser--electron X-ray generator (LEXG). We obtain general relations connecting the polarization state of outgoing X-ray radiation to the polarization state of laser beam as well as to the parameters of electron beam. We demonstrate that finite electron beam emittance causes a partial depolarization of initially fully polarized X-ray radiation even when the laser beam is fully polarized. We demonstrate with a number of numerical experiments that finite electron beam emittance can in some cases fundamentally alter the polarization state of X-ray radiation as compared to the polarization state of X-ray radiation scattered by electron beam with a zero emittance. Possible applications of polarized LEXG's radiation are discussed. 
\end{abstract}


\maketitle

\section{Introduction} 
Thomson scattering of laser beam photons by moderately relativistic electrons (with energy on the order of tens MeV) has been proposed as a source of hard X-ray radiation in the range of 10--100 keV \cite{huang1998laser,carroll1990near,sprangle1992tunable}. The X-ray radiation from such a source (laser--electron X-ray generator -- LEXG) is expected to be tightly collimated and therefore to possess a high spectral brilliance thus making it suitable for a wide range of applications \cite{oliva2010start, jacquet2014high} including the spectroscopy (EXAFS and XANES) \cite{bessonov2009design}, ordinary and phase X-ray tomography \cite{achterhold2013monochromatic, eggl2015x}, X-ray structural analysis \cite{abendroth2010x} as well as in biology and medicine \cite{jacquet2015radiation,bessonov2003laser}.

Theoretical and numerical analysis of properties of the LEXG radiation has attracted significant attention \cite{PhysRevSTHartemann2005, PhysRevSTAB.7.060703, brown2004three, bulyak2005parameters, sun2011theoretical}. However one aspect has not, in our opinion, been sufficiently addressed yet. This is the polarization of the outgoing X-ray radiation and its dependence on the laser beam polarization and the properties of electron beam -- especially on its emittance. Some formulas for the polarization of the Compton back-scattered gamma radiation were presented in \cite{sun2011theoretical} but the authors focused their analysis primarily on the gamma radiation flux angular dependence and its spectral brilliance while giving polarization only a precursory treatment. It should be noted that the polarization of back-scattered gamma radiation is important in nuclear physics and it has been experimentally measured \cite{fukuda2003polarimetry}. In addition, it was also used as a tool to measure the polarization of electron and positron beams \cite{baylac2002first}. 

From general considerations it follows that a finite emittance of electron beam should cause degradation of the polarization degree of X-ray radiation changing it from unity (full polarization), which is expected for the illumination by a fully polarized laser beam. Moreover, all three Stocks parameters of the X-ray radiation can change significantly if the electron beam has a finite emittance. In the present paper we derive general formulas for the polarization of the X-ray radiation emitted by LEXG. Using the specially written program code we conduct a number of numerical experiments for a range of laser and electron beam parameters. We demonstrate how a finite emittance of electron beam degrades and generally affects the X-ray radiation polarization.

\section{Theory of LEXG polarization}
In the case considered in this paper -- of X-ray radiation in the range of tens kiloelectronvolts produced by Thomson scattering of an optical laser beam on an electron beam with a modest electron energy on the order of tens megaelectronvolts -- the following approximations hold: photon--electron scattering can be treated as a classical scattering of zero mass particles (photons) by finite mass particles (electrons). This approximation neglects non-linear, quantum and wave effects as well as the influence of scattering on the electron beam dynamics. The process is characterized by a differential cross-section, which depends in the general case on momentums and energies of both photons and electrons and on the polarization of photons. In the LEXG X-ray radiation is assumed to arise as a result of collisions of trains of electron bunches with synchronized trains of optical (laser) pulses. The electron bunches can circulate in a storage ring or be produced by a linear accelerator.

The polarization of both laser light and X-rays can be characterized by a hermitian polarization matrix of dimension $2\times2$, which can be represented in the following form
\begin{equation}
\mathbf{I}=\frac{I}{2}\left(\begin{array}{cc}
1-\xi_3&\xi_1+i\xi_2\\
\xi_1-i\xi_2&1+\xi_3
\end{array}
\right),
\label{q000}
\end{equation}
where $I$ -- is the radiation intensity and $\xi_1$,  $\xi_2$ è  $\xi_3$ -- are the Stocks parameters ($|\xi_m|\le1$). The polarization degree of the radiation is defined as \cite{landau2004theoretical} \begin{equation}
p=\sqrt{\xi_1^2+\xi_2^2+\xi_3^2}=\sqrt{1-4|\mathbf{I}|/I^2}\le1,
\label{q001}
\end{equation}
where the intensity of the radiation can be expressed as $I=Sp\mathbf{I}$.

It is well known that a general expression for the number of scattering events of a particle with 4D momentum $p_1=(E_1,\mathbf{p}_1)$ by a particle with 4D momentum $p_2=(E_2,\mathbf{p}_2)$ in the direction $\mathbf{n}$ ($|\mathbf{n}|=1$), in the energy interval $dE$ and into a unit of four dimensional volume $dx^4=dr^3dt$ \footnote{We assume that the speed of light $c=1$ in this paper.} can be written as \cite{landau2004theoretical}
\begin{equation}
\frac{dN}{d\Omega dE dx^4 dp_1^3 dp_2^3}=D\sigma_E \sqrt{(j^k_{p_1}j_{k,p_2})^2-j^2_{p_1}j^2_{p_2}},
\label{q00}
\end{equation}
where the summing over index $k$ is implied, $d\Omega$ is a solid angle element in the direction $\mathbf{n}$, $D\sigma_E$ is the differential scattering cross-section of the particle with 4D momentum $p_1$ on the particle with 4D momentum $p_2$ in the energy interval $dE$. The currents of particles $p_1$ and $p_2$ are denoted as $j^k_{p_1}=j^k_{p_1}(\mathbf{r}, \mathbf{p}_1, t)$ and $j^k_{p_2}=j^k_{p_2}(\mathbf{r}, \mathbf{p}_2, t)$, respectively. To take the photon polarization into account the scalar variables $N$ and $D\sigma$ in (\ref{q00}) should be replaced with $2\times2$ matrices similar to one defined in (\ref{q000}).

From expression (\ref{q00}) taking into account the radiation polarization and assuming that particle $1$ represents photons with zero mass it is possible to obtain that the number of photons with momentum $\mathbf{p}_1=\hbar\mathbf{k}$ ($k=\omega_l$) in a fixed polarization state scattered by the electrons with momentum $\mathbf{p}_2=\mathbf{p}=m_e\gamma\mathbf{v}$ in the direction $\mathbf{n}$ from a unit of four dimensional volume $dx^4$ in the frequency interval $d\omega=dE/\hbar$ \cite{PhysRevSTHartemann2005} is
\begin{equation}
\frac{d\mathbf{N}}{d\Omega d\omega dx^4 dp^3 dk^3}=D\boldsymbol{\sigma} j^k_{ph}j_{k,e},
\label{q01}
\end{equation}
where $j^k_{e}=j^k_{ph}(\mathbf{r}, \mathbf{p}, t)=n_e(1,\mathbf{v})$ -- is the four dimensional electron current with momentum $\mathbf{p}$, $j^k_{ph}=j^k_{ph}(\mathbf{r}, \mathbf{k}, t)=n_{ph}(1,\mathbf{n}_l)$ -- is the four dimensional photon current with wave vector $\hbar\mathbf{k}$ (its square is zero), $\gamma=E_e/m_e$ -- is the electron $\gamma$ factor for energy $E_e$, $D\boldsymbol{\sigma}$ -- is the matrix of differential scattering cross-sections of a photon on an electron in frequency interval $d\omega$ and $n_e(\mathbf{r}, t)$ and $n_l(\mathbf{r}, t)$ -- are the spatial densities of the electrons and photons in respective beams. 

The cross-section matrix can be expressed through the polarization matrix of laser radiation  $\boldsymbol{\Xi}$ as follows
\begin{equation}
D\boldsymbol{\sigma}=\frac{d\boldsymbol{\sigma}}{d\Omega d\omega}=\mathbf{M}_{\sigma}^T\boldsymbol{\Xi}\mathbf{M}_{\sigma},
\label{q02}
\end{equation}
where $\mathbf{M}_{\sigma}$ -- is a symmetrical non-hermitian transformation matrix for the radiation field amplitudes, which is defined below, and $\boldsymbol{\Xi}$ is defined as
\begin{equation}
\boldsymbol{\Xi}=\frac{1}{2}\left(\begin{array}{cc}
1-\xi_{L3}&\xi_{L1}+i\xi_{L2}\\
\xi_{L1}-i\xi_{L2}&1+\xi_{L3}
\end{array}\right),
\label{q03}
\end{equation}
where $\xi_{L1}$,  $\xi_{L2}$ è  $\xi_{L3}$ -- are the Stocks parameters of laser radiation.

The number and polarization of emitted photons defined in ({\ref{q01}) should be integrated over electron and laser photon momentums as well as over time to obtain the X-ray radiation flux and polarization (in the matrix form) emitted from an element of the spatial volume
\begin{multline}
\mathbf{I}_{V,\Omega, \omega}(\mathbf{r}, \mathbf{n}, \omega)=\\
\nu\int\limits_{V_{\mathbf{k}}}\int\limits_{V_{\mathbf{p}}}\int\limits_{-\infty}^{+\infty}D\boldsymbol{\sigma} (1-\mathbf{vn}_l)n_e(\mathbf{r}, \mathbf{p}, t)n_{ph}(\mathbf{r}, \mathbf{k}, t)dtdp^3dk^3,
\label{q04}
\end{multline}
where $\nu$ -- is the circulation frequency of electron bunches in the storage ring of LEXG or pulse frequency in the trains produced by the linear accelerator. Expression (\ref{q04}) gives the flux and polarization of X-ray photons averaged over a large number of laser pulses and electron bunches from a 3D spatial volume element $dr^3$ into a solid angle $d\Omega$ and frequency interval $d\omega$.

Using formula (\ref{q04}) the average flux and polarization of X-ray photons in solid angle $d\Omega$ and the frequency interval $d\omega$ can be calculated as
\begin{equation}
\mathbf{I}_{\Omega, \omega}=\int\limits_{V}\mathbf{I}_{V,\Omega, \omega}(\mathbf{r}, \mathbf{n}, \omega)dr^3,
\label{q05}
\end{equation}
whereas the average spectral brilliance of the X-ray radiation in the direction $\mathbf{n}$ and at the frequency $\omega$ is
\begin{equation}
\mathbf{B}_{\mathbf{n}, \omega}=\delta\omega\int\limits_{-\infty}^{+\infty}\mathbf{I}_{V,\Omega, \omega}(\mathbf{r}_0+s\mathbf{n}, \mathbf{n}, \omega)ds,
\label{q06}
\end{equation}
where $\mathbf{r}_0$ -- is position of the observation point and $\delta\omega$ -- is a narrow spectral interval usually assumed to be equal to $10^{-3}\omega$.

Let's now introduce two stationary coordinate frames: $(x,y,z)$ and $(x',y',z')$. The first is related to the electron beam with its z-axis parallel to the average velocity of electron bunches and the second -- to the laser beam with its $z'$-axis parallel to the average direction of laser pulses. For such a choice of coordinate systems the collision angle between the laser and electron beams is close to $\pi$. The general geometry of the electron beam--laser beam interaction used in this paper is shown in Fig.\ref{scheme}.

To calculate the X-ray radiation polarization, as was noted above, a field amplitude transformation matrix $\mathbf{M}_0$ should be used instead of a scalar cross-section (see (\ref{q02})). Such a matrix can be generally expressed as
\begin{equation}
\mathbf{M}_{\sigma}=\mathbf{O}^T\mathbf{M}\mathbf{O},
\label{q101}
\end{equation}
where $O$ -- is an orthogonal matrix representing the rotation by the angle $\chi$, which is the angle between the plane of vectors $\mathbf{v}$ and $\mathbf{n}$ and the plane of axes $y$ and $z$. Diagonal matrix $\mathbf{M}$ is
\begin{equation}
\mathbf{M}=\left(\begin{array}{cc}
\sqrt{m_{11}}&0\\
0&\sqrt{m_{22}}
\end{array}
\right),
\label{q102}
\end{equation}
where $m_{11}$ and $m_{22}$ are scalar scattering cross-sections for s- and p-polarized laser beams, respectively. It can be demonstrated using (\ref{q02}) and (\ref{q101}) that elements of the cross-section matrix $D\boldsymbol{\sigma}$
\begin{equation}
D\boldsymbol{\sigma}=\frac{\sigma_0}{2}\left(\begin{array}{cc}
1-\eta_3&\eta_1+i\eta_2\\
\eta_1-i\eta_2&1+\eta_3
\end{array}\right)
\label{q104}
\end{equation}
can be expressed as
\begin{align}
\sigma_0=&\frac{1}{2}(m_{11}+m_{22})-\nonumber\\
&\frac{1}{2}((\xi_{L3}\cos2\chi +\xi_{L1}\sin2\chi )(m_{11}-m_{22})),\label{q105}\\
\sigma_0\eta_3=&-\frac{1}{2}(m_{11}-m_{22})\cos2\chi+\nonumber\\
&\frac{\xi_{L3}}{2}((m_{11}+m_{22})\cos^{2}2\chi+2\sqrt{m_{11}m_{22}}\sin^{2}2\chi)+\nonumber\\
&\frac{\xi_{L1}}{2}(m_{11}+m_{22}-2\sqrt{m_{11}m_{22}})\sin2\chi\cos2\chi,\label{q106}\\
\sigma_0\eta_1=&-\frac{1}{2}(m_{11}-m_{22})\sin2\chi+\nonumber\\
&\frac{\xi_{L1}}{2}((m_{11}+m_{22})\sin^{2}2\chi+2\sqrt{m_{11}m_{22}}\cos^{2}2\chi)+\nonumber\\
&\frac{\xi_{L3}}{2}(m_{11}+m_{22}-2\sqrt{m_{11}m_{22}})\sin2\chi\cos2\chi,\label{q107}\\
\sigma_0\eta_2=&\xi_{L2}\sqrt{m_{11}m_{22}}.\label{q108}
\end{align}
\begin{figure}
\includegraphics[width=1\linewidth]{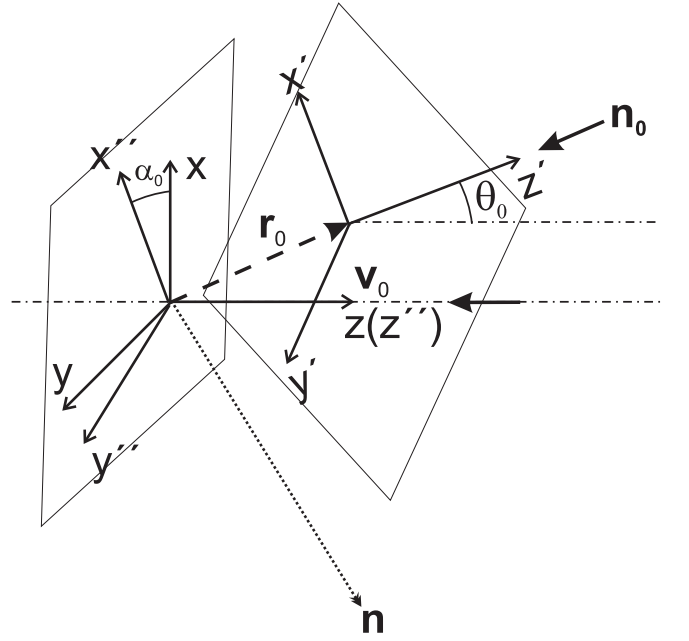}
\caption{The scheme of electron beam -- laser beam interaction in LEXG. Three coordinate systems used (see (\ref{q181})--(\ref{q186})) as well as angles, vectors and shifts are shown.}
\label{scheme}
\end{figure}
From expressions (\ref{q02})--(\ref{q03}) and (\ref{q101})--(\ref{q102}) it follows that the polarization degree of X-ray radiation (\ref{q001}) (assuming that the angular spread of electron momentums in the electron beam is zero) is transformed as 
\begin{equation}
p=\sqrt{1-\frac{m_{11}m_{22}}{I^2}(1-p_L^2)},
\label{q109}
\end{equation}
where $p_L=\sqrt{\xi_{1L}^2+\xi_{2L}^2+\xi_{3L}^2}$ and intensity $I=Sp\mathbf{I}_{V,\Omega,\omega}$. The value of $I$ is equal to the X-ray radiation intensity produced by scattering of a naturally polarized laser beam. In particular, from (\ref{q109}) it can be seen that the fully polarized laser radiation ($p_L=1$) produces the fully polarized X-ray radiation if the angular spread of electron momentums in the electron beam is neglected. This assumption is in some sense equivalent to assuming that the electron beam transversal emittance is zero and fixing the electron beam waist width since it also depends on the emittance. Below when we speak about zero electron beam emittance we will understand it in this sense. (\ref{q105})--(\ref{q108}) correspond to formulas (35) from \cite{sun2011theoretical}. 

Coefficients $m_{11}$ and $m_{22}$ can be obtained from the differential cross-sections for linearly polarized light. For our purpose we can neglect all corrections to this cross-section related to collision angle $\varphi$ between the laser and electron beams. Such corrections are proportional to its difference $\sim\delta\varphi$ from $\pi$ squared: $\delta\varphi^2=2(\mathbf{vn}_l+1)\ll0$ and therefore they are very small. This allows one to use the cross-section for a head-on collision of an electron with a photon when $\mathbf{vn}_l=-1$. As it was shown in \cite{PhysRevSTAB.7.060703}, the differential cross-section for linearly polarized laser beam can be written as 
\begin{align}
\frac{d\sigma}{d\Omega d\omega}=&\frac{3}{8\pi}\frac{\sigma_T}{\gamma^2}\delta(\omega-\omega')\frac{1}{(1-v\cos\theta)^2}\times\nonumber\\
&\left[1-\frac{\sin^2\theta\sin^2\alpha}{\gamma^2(1-v\cos\theta)^2}\right],\label{q061}\\
\frac{\omega'}{\omega_l}&=\frac{1+v}{1-v\cos\theta},\quad v=|\mathbf{v}|,\quad \mathbf{vn}=v\cos\theta, \label{q062}
\end{align}
where frequency $\omega'$ is related to the laser frequency by formula (\ref{q062}), $\alpha$ -- is the angle between the electric field vector of laser beam and axis $y$,  $\sigma_T=6.65\times10^{-25}\;\mbox{cm}^2$ -- is the Thomson cross-section.

Coefficients $m_{11}$ and $m_{22}$ are obtained from (\ref{q061}) when $\alpha=\pi/2$ and $\alpha=0$, respectively.

Let's consider the spatial distributions of electrons and photons. Since the betatron and synchrotron oscillations of electrons in a storage ring can be considered approximately independent from each other \cite{wiedemann2015particle} the volume densities $n_e$ and $n_{ph}$ can be factorized into a transversal part ($\perp$) and a longitudinal part ($\parallel$), which determine the distribution of electrons and photons by the transversal momentum and coordinate and by the energy and longitudinal coordinate, respectively. So, the electron and photon distributions can be written as
\begin{align}
n_e&=N_ef_{e\parallel}(z-z_0-t,\gamma)f_{e\perp}(\boldsymbol{\rho}-\boldsymbol{\rho}_0,\mathbf{p}_\perp),\label{q111}\\
n_{ph}&=N_{ph}f_{ph\parallel}(z'-t-\Delta t,\omega_l)f_{ph\perp}(\boldsymbol{\rho}',\mathbf{k}_\perp),\label{q112}
\end{align}
where $N_e$ è $N_{ph}$ -- are the total number of electrons and photons in an electron bunch and a laser pulse, respectively, $\Delta t$ -- is the temporal delay between an electron bunch and a laser pulse and  $\mathbf{r}_0=(\boldsymbol{\rho}_0, z_0)$ -- is a vector accounting for the spatial shift between an electron bunch and a laser pulse.

The longitudinal component of photon distribution function in (\ref{q112}) is assumed to have a Gaussian form in space
\begin{equation}
f_{ph\parallel}(z',\omega_l)=\frac{1}{\sqrt{\pi}\tau}\exp\left\{-\left(\frac{z'}{\tau}\right)^2\right\}\delta(\omega_l-\omega_0),
\label{q12}
\end{equation}
where $\tau$ -- is the laser pulse length and $\omega_0$ -- is the average frequency of laser radiation. In formula (\ref{q12}) the laser pulse is considered to be monochromatic, that is usually justified for sufficiently long ($>0.5$ ps) pulses in the case of LEXG.

The longitudinal component of electron distribution function in (\ref{q111}) by coordinate $z$ and the energy ($\gamma$) is also Gaussian
\begin{equation}
f_{e\parallel}(z,\gamma)=\frac{1}{\pi l_e\Delta\gamma}\exp\left\{-\left(\frac{z}{l_e}\right)^2\right\}\exp\left\{-\left(\frac{\gamma-\gamma_0}{\Delta\gamma}\right)^2\right\},
\label{q13}
\end{equation}
where $l_e$ -- is the electron bunch length, $\gamma_0$ -- is the average gamma factor of electron bunch and $\Delta\gamma$ -- is the gamma factor distribution width. 

For transversal distribution functions in (\ref{q111}) and (\ref{q112}) the common form used in literature \cite{wiedemann2015particle} is that of a Gaussian beam
\begin{align}
f_{e\perp}&(\boldsymbol{\rho},\mathbf{p}_\perp)=\frac{1}{\pi^2 w_{e,x}(z-z_0)w_{e,y}(z-z_0)\Delta p_{\perp,x}p_{\perp,y}}\times\nonumber\\
&\exp\left\{-\left(\frac{x}{w_{e,x}(z-z_0)}\right)^2-\left(\frac{y}{w_{e,y}(z-z_0)}\right)^2\right\}\times\nonumber\\
&\exp\left\{-\left(\frac{p_{\perp,x}}{\Delta p_{\perp,x}}\right)^2-\left(\frac{p_{\perp,y}}{\Delta p_{\perp,y}}\right)^2\right\},\label{q151}\\
f_{ph\perp}&(\boldsymbol{\rho}',\mathbf{n}_l)=\nonumber\\
&\frac{1}{\pi w_{ph}^2(z')}\exp\left\{-\left(\frac{\boldsymbol{\rho}'}{w_{ph}(z')}\right)^2\right\}\delta(\mathbf{n}_l-\mathbf{n}_0),\label{q152}
\end{align}
where $\Delta p_{\perp,x}$ and $\Delta p_{\perp,y}$ -- are the transversal electron momentum spreads along $x$ and $y$ axes, respectively and $\mathbf{n}_0$ -- is a unit vector in the direction of laser beam. It is well known \cite{wiedemann2015particle} that
\begin{align}
\Delta p_{\perp,x}&=\left(\frac{\varepsilon_x}{\beta_x\gamma}\right)^{1/2},\label{153}\\
\Delta p_{\perp,y}&=\left(\frac{\varepsilon_y}{\beta_y\gamma}\right)^{1/2},\label{154}
\end{align}
and functions $w_{e,x}$, $w_{e,x}$ and $w_{ph}$ are \cite{wiedemann2015particle}
\begin{align}
w_{e,x(y)}^2(z)&=\frac{\varepsilon_{x(y)}}{\gamma}\left(\beta_{x(y)}+\frac{z^2}{\beta_{x(y)}}\right),\label{q161}\\
w_{ph}^2(z')&=\frac{1}{2\omega_l}\left(l_r+\frac{z'^2}{l_r}\right),\label{q162}
\end{align}
where $\beta_{x(y)}$ -- is the beta function of electron beam at the interaction point in direction of $x$($y$) axis, $\varepsilon_{x(y)}$ -- is the transversal emittance of electron beam in direction of $x$($y$) axis and $l_r$ -- is the Rayleigh length of laser beam. In (\ref{q152}) the photon transversal momentum distribution is in the form of a delta function, which corresponds to the approximation considered above where the scattering cross-section does not depend on the collision angle between an electron and a photon. 

In (\ref{q13}), (\ref{q151}) and (\ref{q152}) the electron energy distribution as well as the electron transversal momentum distribution are considered to be independent from the spatial coordinates -- an assumption, which is justified in case of LEXG because the transversal sizes of both beams are relatively large.

One can note that time is only present in the longitudinal components of electron and photon densities $f_{e\parallel}$ and $f_{ph\parallel}$. Therefore the integral over time in (\ref{q04}) can be computed analytically
\begin{multline}
\int\limits_{-\infty}^{+\infty}f_{e\parallel}(z-z_0-t,\gamma)f_{ph\parallel}(z'-t-\Delta t,\omega_l)dt=\\
\frac{1}{\pi\Delta\gamma\sqrt{l_e^2+\tau^2}}\exp\left\{-\frac{(z-z'-z_0+\Delta t)^2}{l_e^2+\tau^2}\right\}\times\\
\exp\left\{-\left(\frac{\gamma-\gamma_0}{\Delta\gamma}\right)^2\right\}\delta(\omega_l-\omega_0).
\label{q14}
\end{multline}

We are now ready to write the final expression for polarization matrix $\mathbf{I}_{V,\Omega, \omega}(\mathbf{r}, \mathbf{n}, \omega)$ from (\ref{q04}) using the cross-sections and distributions obtained above. So, taking into account integral (\ref{q14}) and averaging the cross-section matrix (\ref{q104}) over $\gamma$, expression (\ref{q04}) can be transformed into
\begin{widetext}
\begin{multline}
\mathbf{I}_{V,\Omega, \omega}(\mathbf{r}, \mathbf{n}, \omega)=\frac{2\nu N_eN_{ph}}{\pi^{7/2}\sqrt{l_e^2+\tau^2}\Delta p_{\perp,x}\Delta p_{\perp,y}w_{e,x}(z-z_0)w_{e,y}(z-z_0)w_{ph}^2(z')}\times\\
\exp\left\{-\frac{(z-z'-z_0+\Delta t)^2}{l_e^2+\tau^2}-\left(\frac{x-x_0}{w_{e,x}(z-z_0)}\right)^2-\left(\frac{y-y_0}{w_{e,y}(z-z_0)}\right)^2-\left(\frac{\boldsymbol{\rho'}}{w_{ph}(z')}\right)^2\right\}\times\\
\int\limits_{V_{\mathbf{p}_\perp}} D\boldsymbol{\sigma}_{\gamma}(\mathbf{nv})\exp\left\{-\left(\frac{p_{\perp,x}}{\Delta p_{\perp,x}}\right)^2-\left(\frac{p_{\perp,y}}{\Delta p_{\perp,y}}\right)^2\right\}dp_\perp^2,
\label{q19}
\end{multline}
\end{widetext}
where $D\boldsymbol{\sigma}_\gamma$ -- is the cross-section polarization matrix averaged over the distribution of $\gamma$, coordinates $(x,y,z)$ and $(x',y',z')$ are related to each other by the following expressions (see Fig.\ref{scheme}) consisting of two sequential rotations around $x$ and $z$ axes (taking into account the axial symmetry of laser beam)
\begin{align}
z'&=\cos\theta_0z''-\sin\theta_0y'',\label{q181}\\
y'&=\sin\theta_0z''+\cos\theta_0y'',\label{q182}\\
x'&=x'',\label{q183}\\
z''&=z,\label{q184}\\
y''&=\cos\alpha_0 x-\sin\alpha_0 y,\label{q185}\\
x''&=\sin\alpha_0 x+\cos\alpha_0 y.\label{q186}
\end{align}
Here $\theta$ is the observation angle and angle $\alpha_0$ corresponds to the rotation around axis $z$. In formulas (\ref{q181}) and (\ref{q182}) the collision angle $\theta_0$ ($\cos\theta_0=\mathbf{v}_0\mathbf{n}_0$) is not assumed to be equal to $\pi$ as in differential cross-section matrix (\ref{q104}). 
\begin{table}[t]
	\centering
		\begin{tabular}{|r|c|l|}
		\hline
		\textbf{Parameter}&\textbf{Value}&\textbf{Notes}\\
		\hline
		$E_e$& 35--50 MeV & Electron energy\\
		\hline
		$\gamma$& 70--100 & Gamma factor \\
		\hline
		$N_e$& $0.6\times10^{10}$ &Bunch charge 1 nQ \\
		\hline
		$\Delta\gamma/\gamma$& $10^{-2}$ & Relative electron energy spread\\	
		\hline
		$E_L$& 20 $\mu$J & Laser pulse energy\\	
		\hline
		$\hbar\omega_L$& 1.1 eV & Laser photon energy\\
		\hline
		$\nu$& 79 MHz & Frequency of laser pulses\\
		\hline
		$\tau$ & 30 ps & Electron bunch/laser pulse length \\
		\hline
		$\hbar\omega_X$& 44 keV & X-ray photon energy\\		
		\hline
		$\varepsilon_{x}, \varepsilon_{y}$& 5 $\mbox{mm}\times\mbox{mrad}$& Normalized emittances\\		
		\hline
		$\beta_x$, $\beta_y$& 10 mm & Beta function\\		
		\hline
		$l_r$& 5.4 mm & Rayleigh length\\		
		\hline
		\end{tabular}
	\caption{The LEXG parameters used for numerical experiments}
	\label{tab1}
\end{table}
Taking into account that cross-section (\ref{q062}) contains a delta-function, the averaging over $\gamma$ in $D\boldsymbol{\sigma}$ (see (\ref{q105})--(\ref{q108})) can be performed analytically. Then the averaged (over $\gamma$) diagonal matrix elements $\overline{m_{11}}$, $\overline{m_{22}}$ and  $\overline{m_{12}}=\overline{\sqrt{m_{11}m_{22}}}$ can be obtained
\begin{align}
\overline{m_{11}}&=\frac{3\sigma_T}{64\pi\sqrt{\pi}\Delta\gamma\omega_l}\sqrt{\frac{\omega}{\omega_l}}
\frac{1}{\displaystyle \sqrt{1-\frac{\omega}{\omega^2}{4}}}e^{-K},\label{q24}\\
\overline{m_{22}}&=m_{11}\left(1-\frac{\omega}{\omega_l}\frac{\theta^2}{2}\right)^2,\label{q25}\\
K&=\frac{\displaystyle \left[\sqrt{\frac{\omega}{\omega_l}\frac{1}{1-\omega\theta^2/(4\omega_l)}}-2\gamma_0\right]^2}{\displaystyle 4\Delta\gamma^2},\label{q251}\\
\overline{m_{12}}&=\sqrt{\overline{m_{11}}\,\overline{m_{22}}},\label{q252}\\
\theta^2&=2(1-\mathbf{vn}).\nonumber
\end{align}
\begin{figure}
\includegraphics[width=1\linewidth]{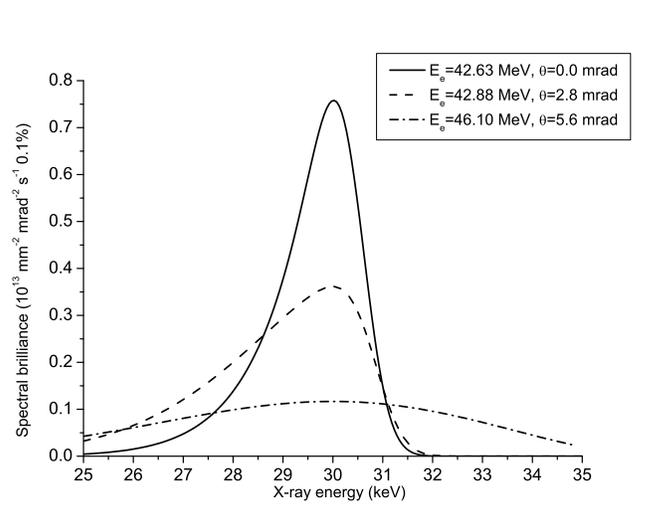}
\caption{The X-ray spectral brilliance for three observation angles $\theta$ and the corresponding electron beam energies. The angles and energies were chosen to produce the  radiation peaks at the same energy of 30 keV.}
\label{f1}
\end{figure}
As it is clear from (\ref{q19}), to calculate the X-ray polarization one should average the polarization matrix $D\boldsymbol{\sigma}_\gamma$ from (\ref{q104}) over the transversal spread of electron momentums. This operation is equivalent to averaging of expressions (\ref{q105})--(\ref{q108}) and in a general case leads to a partial depolarization of initially fully polarized X-ray radiation. The averaging can not be performed analytically for an arbitrary observation angle. However, in the special case of zero observation angle $\theta$ the matrix elements $m_{11}$ and $m_{22}$ does not depend on the angle $\chi$, which in this case is equal to the azimuth angle between an electron velocity and the average velocity of electron bunches. Now expressions (\ref{q105})--(\ref{q108}) can be analytically averaged over angle $\chi$, which results in the following expressions
\begin{align}
\overline{\sigma_0}=&\frac{1}{2}(\overline{m_{11}}+\overline{m_{22}}),\label{q26}\\
\overline{\sigma_0\eta_3}=&\frac{\xi_{L3}}{4}(\sqrt{\overline{m_{11}}}+\sqrt{\overline{m_{22}}})^2,\label{q27}\\
\overline{\sigma_0\eta_1}=&\frac{\xi_{L1}}{4}(\sqrt{\overline{m_{11}}}+\sqrt{\overline{m_{22}}})^2,\label{q28}\\
\overline{\sigma_0\eta_2}=&\xi_{L2}\sqrt{\overline{m_{11}}\,\overline{m_{22}}}.\label{q29}
\end{align}
To obtain polarization matrix $\mathbf{I}_{V,\Omega,\omega}$ expressions (\ref{q26})--(\ref{q29}) need to be further averaged over the angle between the velocity of an electron and the average velocity of electron bunch. The result of such an averaging depends on the ratio of angular width of X-ray distribution in $\overline{m_{11}}$ and $\overline{m_{22}}$ to the width of angular spread of the electron momentums (\ref{q151}). It can be noted that formulas (\ref{q26})--(\ref{q29}) are the same as formulas (38) in \cite{sun2011theoretical}.
\begin{figure}
\includegraphics[width=1\linewidth]{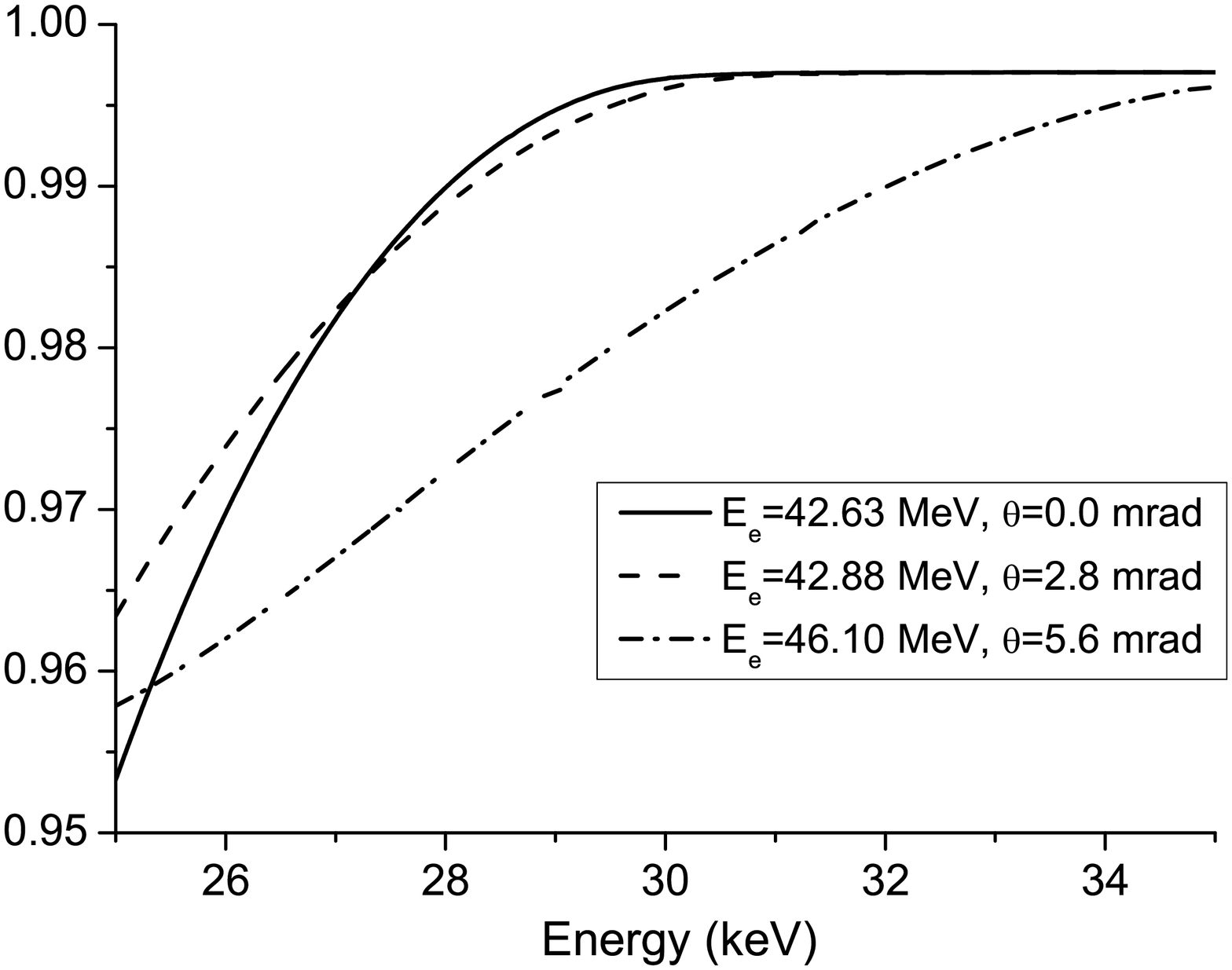}
\caption{The X-ray radiation polarization degree for three observation angles $\theta$ and the corresponding electron energies as function of the X-ray radiation energy. The angles and energies were chosen to produce the radiation peaks at the same energy of 30 keV, as in Fig.\ref{f1}.}
\label{f2}
\end{figure}

\section{Numerical experiments}
To calculate the X-ray radiation characteristics of LEXG a program code was written in the JAVA programming language \cite{TSource}. The code is based on the model outlined in this paper and can compute the full polarization matrix of X-ray radiation for arbitrary LEXG parameters. The code is capable of generating ray sets for a ray-tracing program code like SHADOW \cite{sanchez2011shadow3}for simulation of the X-ray propagation through complex optical systems (especially X-ray beamlines).
 
\begin{figure}
\includegraphics[width=1\linewidth]{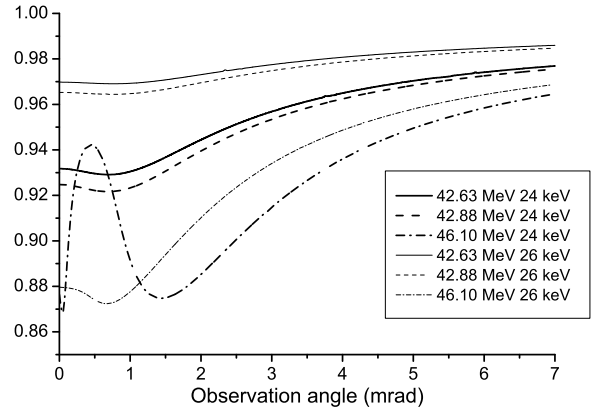}
\caption{The X-ray radiation polarization degree for the three electron beam energies (and corresponding observation angles) from Fig.\ref{f2} and two X-ray energies as function of the observation angle.}
\label{f3}
\end{figure}
To demonstrate how the transversal emittance and other beam parameters influence the polarization of the X-ray radiation we conducted a number of numerical experiments. The main parameters of LEXG used in the calculations are shown in Table \ref{tab1}. These parameters chosen are rather typical for X-ray Thomson sources. The laser pulse energy is supposed to be amplified in a high finesse cavity by about $10^3$ resulting in 20 mJ pulses interacting with electron bunches. The peak laser intensity at  the point of interaction is about $5\times10^{13}$ $W/cm^{2}$ -- sufficiently low to disregard any non-linear effects.

In the figures below the spectral brilliance and polarization of X-ray radiation of LEXG with parameters from Table \ref{tab1} are depicted as functions of various parameters.
\begin{figure}
\includegraphics[width=1\linewidth]{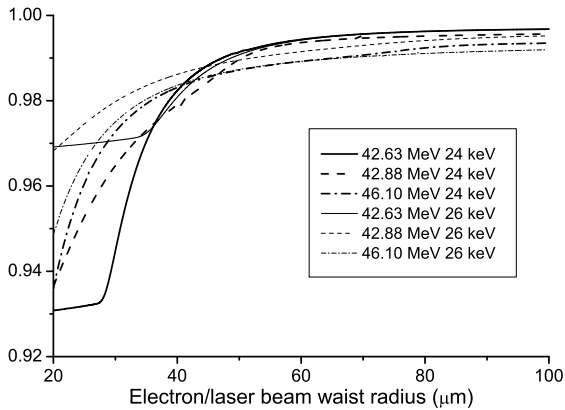}
\caption{The X-ray radiation polarization degree for the three electron beam energies (and corresponding observation angles) from Fig.\ref{f2} and two X-ray energies as function of the electron and laser beam waist radii, which were considered equal. The waist radii were varied by adjusting the beta-function and Rayleigh length.}
\label{f4}
\end{figure}
\begin{figure}
\begin{subfigure}{.5\textwidth}
 \includegraphics[width=1\linewidth]{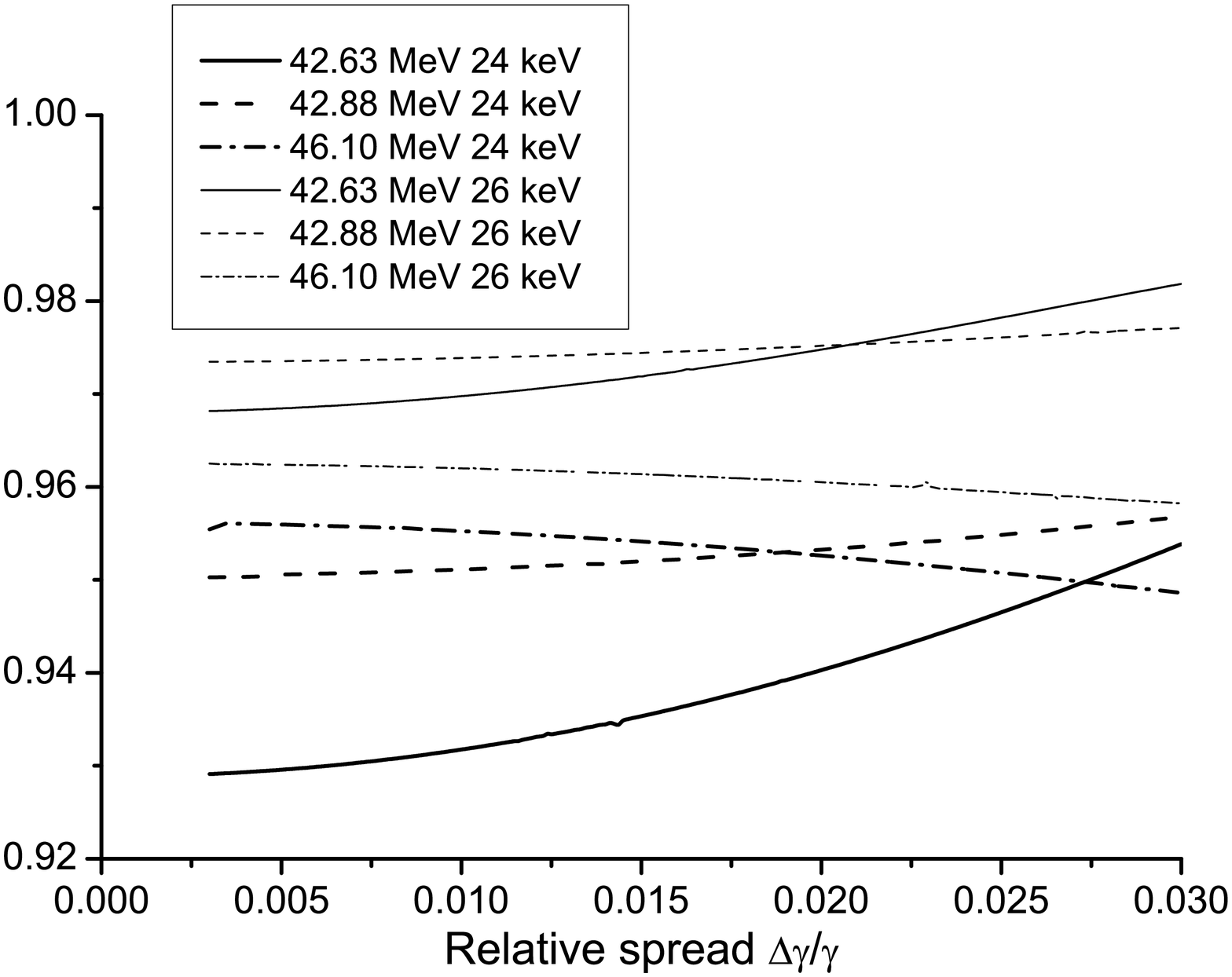}
\end{subfigure}
\begin{subfigure}{.5\textwidth}
 \includegraphics[width=1\linewidth]{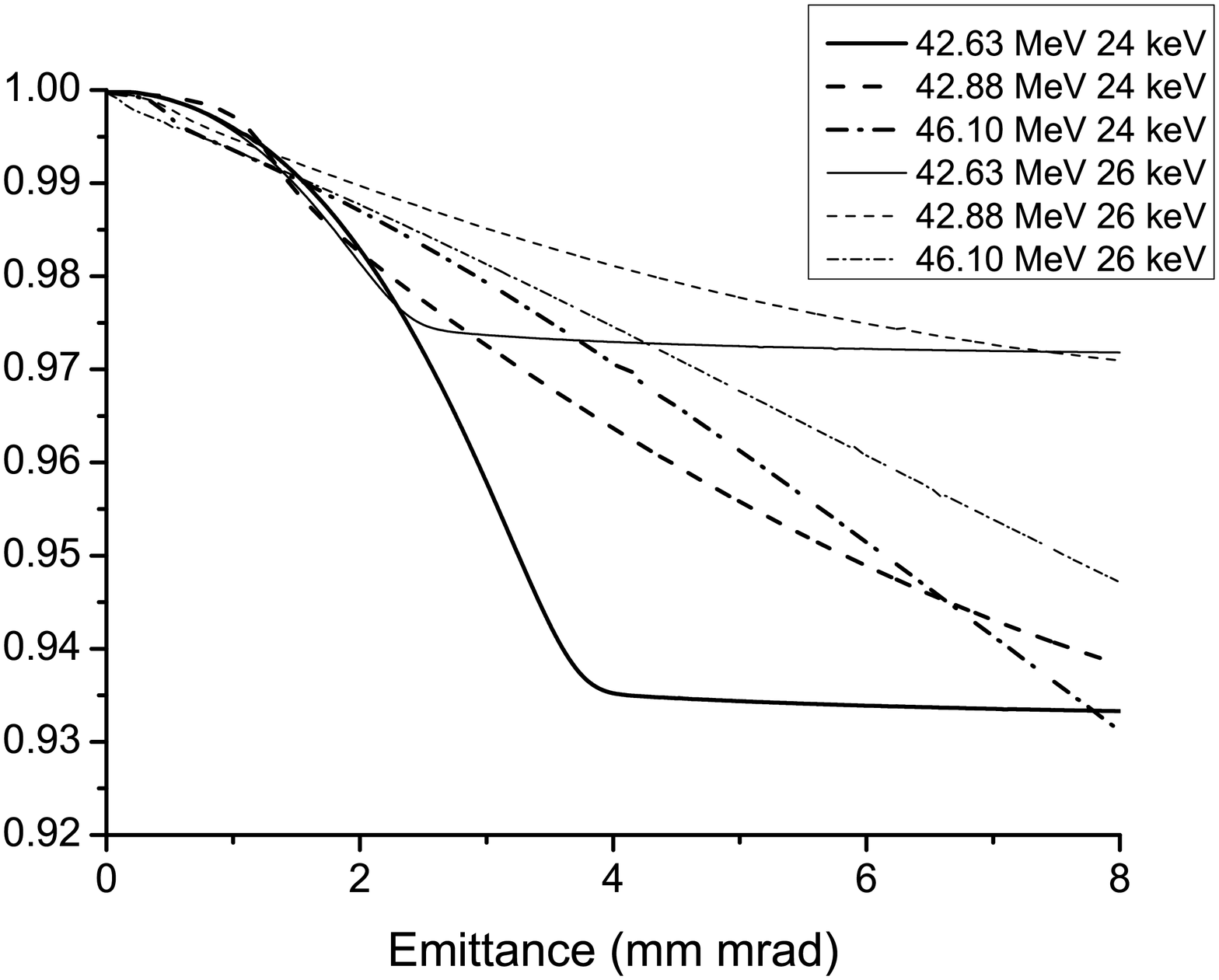}
\end{subfigure}
\caption{The X-ray radiation polarization degree for three electron beam energies (and corresponding observation angles) and two X-ray energies from Fig.\ref{f2} as function of the relative electron energy spread $\Delta\gamma/\gamma$ (top) or electron beam emittance (bottom).}
\label{f5}
\end{figure}
In Fig.\ref{f1} the spectral brilliance of LEXG radiation for a naturally polarized laser beam ($\xi_1=\xi_2=\xi_3=0$) is shown for three different observation angles $\theta$ and electron beam energies $E_e$. The collision angle of the electron and laser beams $\theta_0$ is set at 50 mrad \footnote{In this section the collision angle was redefined as $\pi-\theta_0$ for convenience.}. As it was noted above, the intensity of the naturally polarized light serves as the denominator in the calculation of the Stocks parameters and degree of polarization of the X-ray radiation.

From Fig.\ref{f1} one can see that, in particular, the width of spectral brilliance curves significantly increases and the peak spectral brilliance significantly decreases as the observation angle grows. The curves are asymmetrical with respect to their maxima with long exponential tails in the direction of lower X-ray energies. They become progressively more symmetrical as the observation angle is increased. This asymmetry arises because at zero observation angle the X-ray radiation of higher energies is emitted by higher energy electrons moving strictly in the direction of the electron beam. There are only a few electrons due to the relatively narrow electron energy spectrum of (\ref{q13}). The long exponential tail is created mainly by the electrons moving non-collinearly to the electron beam direction. The number of such electrons decreases exponentially as function of the angle according to (\ref{q151}). 

In Fig.\ref{f2} the X-ray radiation polarization degree for three observation angles $\theta$ and the electron energies $E_e$ as function of the X-ray radiation energy are shown. The electron beam emittance is important here as the polarization degree would be unity without it. The collision angle $\theta_0$ is 50 mrad and the laser beam is linearly polarized with $\xi_1=\xi_3=1/\sqrt{2}$ and $\xi_2=0$ (the polarization plane was rotated by $\pi/4$ around $z$ axis relative to $y$ axis). 

It can be seen that a finite electron beam emittance causes a depolarization of the emitted X-ray radiation, which increases with the observation angle. It is clear that for the parameters listed in Table \ref{tab1} depolarization of the X-ray radiation by a finite electron beam emittance is rather small -- a few percent only. The polarization degree also approaches unity as the X-ray energy grows. Significant depolarization is observed only for the X-ray energy less than its peak value in Fig.\ref{f1}. This can be explained by the fact that the higher energy X-rays arise as a result of scattering by higher energy electrons, which move strictly in the direction of the electron beam, while the lower energy X-rays are produced by electrons moving at some angles to the electron beam direction.

In Fig.\ref{f3}--\ref{f5}} the degree of polarization for the same three electron energies as in Fig.\ref{f2} is plotted against the observation angle, beam waist diameter, relative electron energy spread $\Delta\gamma/\gamma$ and electron beam emittance for two X-ray energies 24 and 26 keV. The latter energies are less than the peak energies in Fig.\ref{f1}. In Fig.\ref{f4}--\ref{f5} observation angles are the same as in Fig.\ref{f2} for the same electron beam energies. Other parameters are also the same as in Fig.\ref{f2}. It can be seen that the polarization degree can exhibit rather complex behavior exhibiting local minima and maxima, see Fig.\ref{f3}. The polarization degree at 26 keV is generally higher than at 24 keV as in Fig.\ref{f2}. The origin of the non-monotonic behavior can be traced back to the interaction between a finite emittance of electron beam with the polarization parameters in (\ref{q105})--(\ref{q108}) during the averaging over electron momentum spread. 

In Fig.\ref{f4} the polarization degree looks like a shelf when the beam's waist is small and the observation angle is zero (solid curves). This behavior is explained by sort of a saturation of the polarization degradation when the transversal electron momentum spread (inversely proportional to the waist radious when the emittance is fixed) exceeds the angle corresponding to the observed X-ray energy according to (\ref{q062}). The saturated polarization degree in this case is determined by average parameters (\ref{q26})--(\ref{q29}) taken at this angle. 

In Fig.\ref{f5} (top) the polarization degree increases with $\Delta\gamma/\gamma$ at smaller observation angles and weakly decreases at larger observation angles. Such a behavior is explained by the fact that at smaller angles as the energy spread increases relatively more X-rays are emitted by electrons moving in the direction of electron beam, which reduces the angular spread of the contributing electrons. At larger angles the behavior switches in the opposite direction because a larger contribution from the electrons moving in the direction of electron beam means a larger angular spread of the contributing electrons. On the other hand as from Fig.\ref{f5} (bottom) the polarization degree initially monotonically decreases as the emittance increases and then saturates by the same reason as in Fig.\ref{f4}.

\begin{figure}
\begin{subfigure}{.5\textwidth}
 \includegraphics[width=1\linewidth]{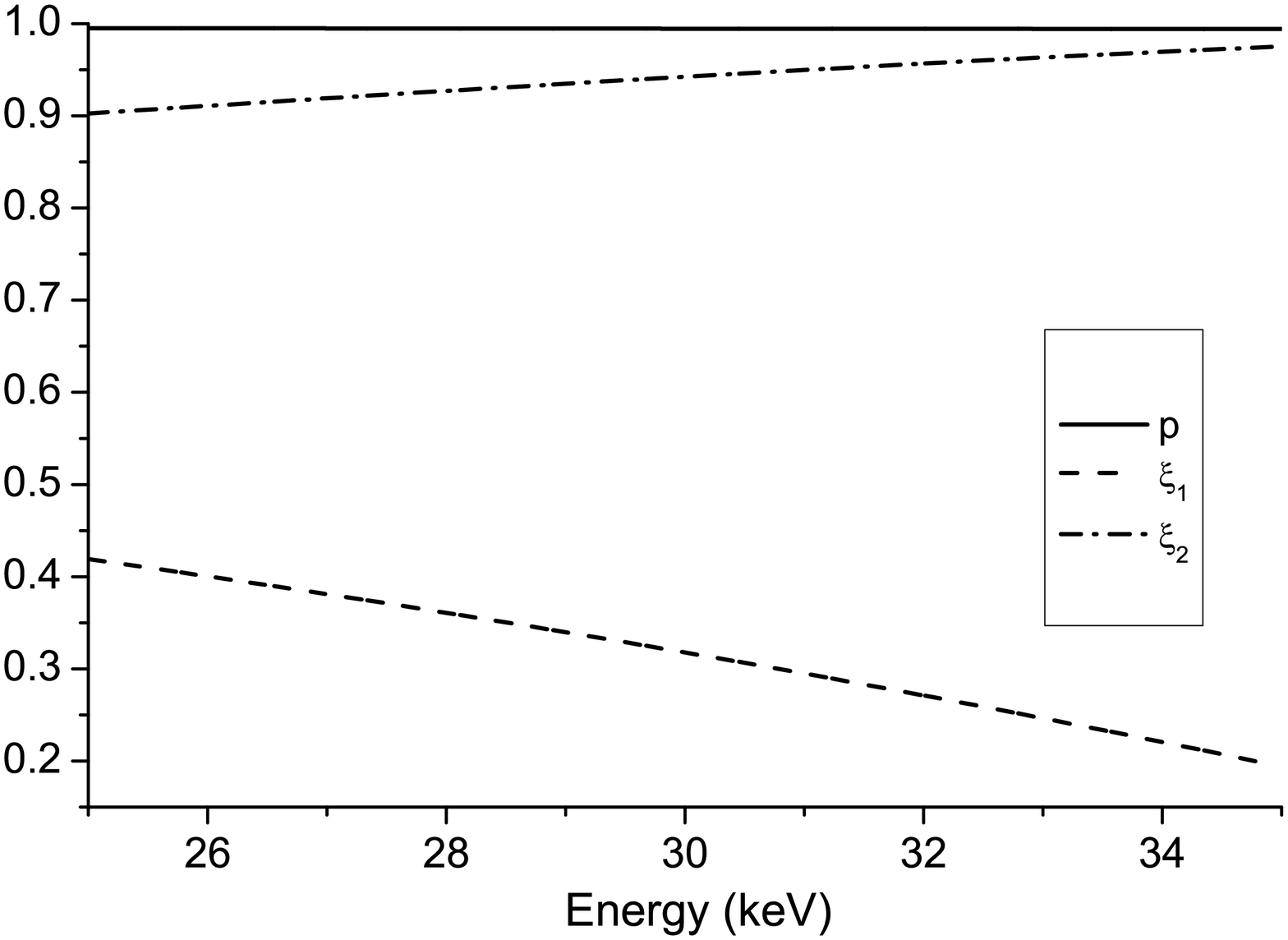}
\end{subfigure}
\begin{subfigure}{.5\textwidth}
 \includegraphics[width=1\linewidth]{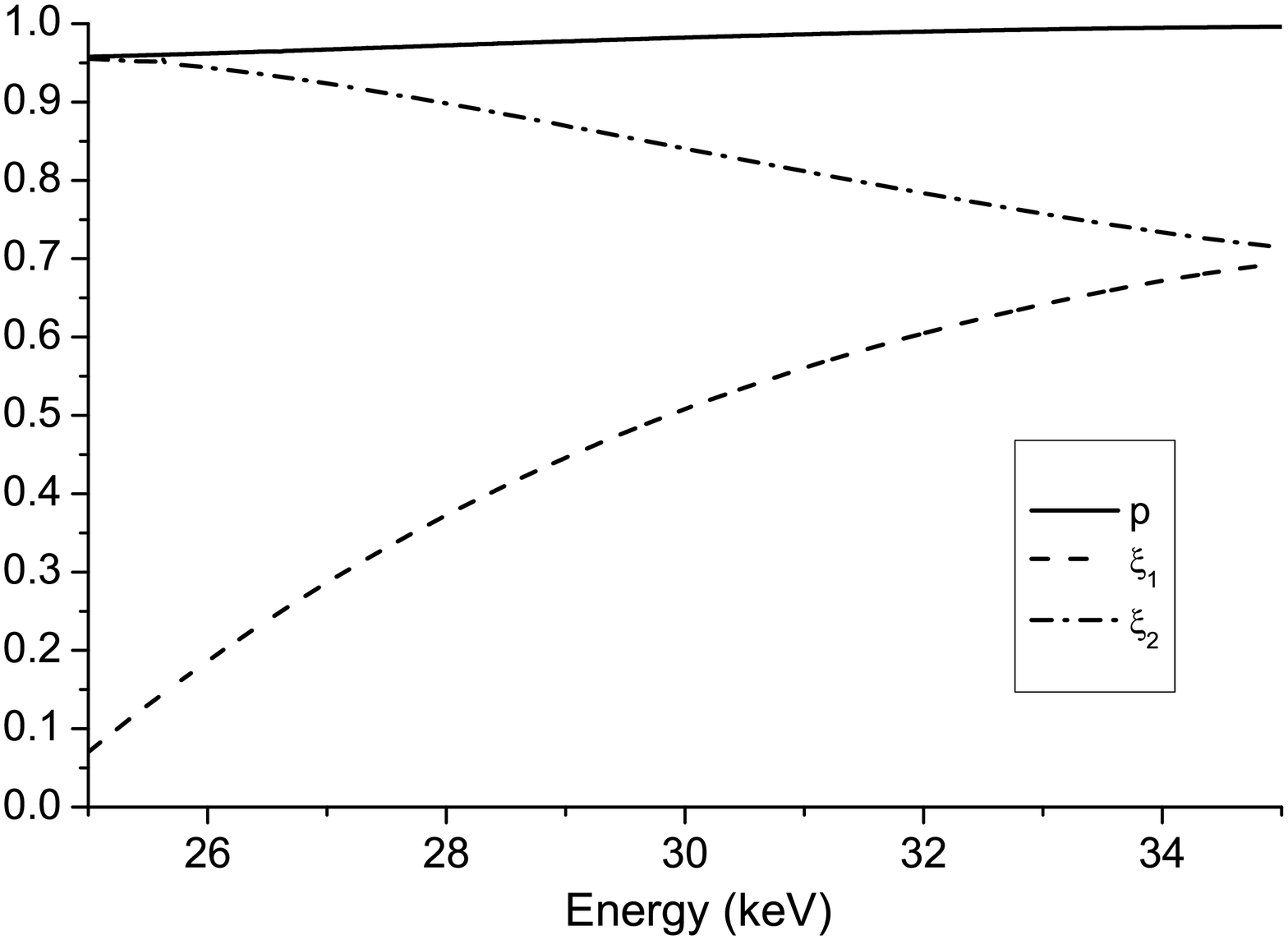}
\end{subfigure}
\caption{The Stocks parameters $\xi_1$ and $\xi_3$ as well as the polarization degree as function of the X-ray energy. The observation angle was 5.6 mrad. The electron beam emittance was zero (top) or not zero (bottom).}
\label{f6}
\end{figure}

In Fig.\ref{f6} Stocks parameters as function of the X-ray energy are shown for observation angle of 5.6 mrad. The electron energy is set at 46.10 MeV. The remaining parameters are the same as in Fig.\ref{f2} for $\theta_0=5.6$ mrad.
From Fig.\ref{f6} it is evident that a finite electron beam emittance radically changes the X-ray polarization. In the top graph X-ray radiation is scattered only by the electrons with momentums parallel to the electron beam direction whereas in the bottom graph higher energy X-ray radiation is scattered mainly by electrons moving in the direction of observer. This makes the higher energy X-rays polarization in the bottom graph the same as for zero observation angle (not shown) whereas in the top graph its polarization remains rotated by about $\pi/4$ as for lower energy X-rays.

In the examples shown above characteristics of LEXG X-ray radiation were considered at a fixed X-ray energy and observation angle. However X-ray beams passing through real optical systems such as synchrotron beamlines often comprise a range of X-ray energies and angles. In other words they need to be appropriately collimated and monochromatized. The polarization state of such X-ray beams can be calculated using a suitable ray-tracing code such as program complex SHADOW mentioned above \cite{sanchez2011shadow3}. SHADOW can accurately trace rays having an arbitrary polarization state. The program code that we used in this article can generate initial ray sets for it (as we said at the beginning of this section), so the performance of real X-ray optical system can be analyzed when using LEXG as a source.  

\section{Conclusion}
In the this paper a working model for the calculation of polarization of X-ray radiation produced by a laser-electron X-ray generator is presented. It allows one to simulate the full polarization state of X-ray radiation for an arbitrary initial polarization state of laser beam. The model takes into account the transversal emittance of electron beam, which has significant influence over the final polarization state of X-ray radiation. The model has been implemented in a specially written program code in JAVA programming language. 

Using a number of numerical experiments we demonstrated that the electron beam emittance causes partial depolarization of the scatted X-ray radiation. However for typical LEXG parameters this depolarization was rather small -- a few percent only. In addition, the orientation of the X-ray radiation polarization can change dramatically when the electron beam emittance is taken into account. 

The model of the X-ray radiation polarization of LEXG presented in this work may be useful for evaluation of the LEXG applications where polarization of the X-ray radiation is important. Such applications include, for example, the study of surface characteristics of the materials by the X-ray ellipsometry \cite{bracco2013surface} and the polarization sensitive spectroscopy -- for instance, investigation of X-ray magnetic circular dichroism near absorption edges in magnetic materials \cite{van2013applications}.

\begin{acknowledgments}
The authors of the paper want to thank Prof. V.I Shvedunov for fruitful discussions concerning the physics of electron storage rings and injectors. This work was supported by the Program of fundamental investigations of the Russian Academy of Science Presidium “Fundamental and applied problems of photonics and physics of new optical materials”.

\end{acknowledgments}

\end{document}